\documentclass{aa}
\usepackage{graphicx,natbib}
\bibpunct{(}{)}{;}{a}{}{,}

\newcommand{\beq}{\begin{equation}}
\newcommand{\eeq}{\end{equation}}
\newcommand{\beqa}{\begin{eqnarray}}
\newcommand{\eeqa}{\end{eqnarray}}

\def\nh{N_{\rm H}}

\def\lhx{L_{\rm hx}}

\def\lmin{L_{\rm min}}
\def\lmax{L_{\rm max}}

\begin{document}

\title{Cumulative hard X-ray spectrum of local AGN: a link to
the cosmic X-ray background}

\author{S.~Sazonov\inst{1,2} \and R.~Krivonos\inst{2,1} \and 
M.~Revnivtsev\inst{1,2} \and E.~Churazov\inst{1,2} \and R.~Sunyaev\inst{1,2}}

\offprints{sazonov@mpa-garching.mpg.de}

\institute{Max-Planck-Institut f\"ur Astrophysik,
           Karl-Schwarzschild-Str. 1, D-85740 Garching bei M\"unchen,
           Germany
     \and   
           Space Research Institute, Russian Academy of Sciences,
           Profsoyuznaya 84/32, 117997 Moscow, Russia
}
\date{Received / Accepted}

\authorrunning{Authors}
%\titlerunning{Cumulative hard X-ray spectrum of local AGE}

\abstract
{}
{We determine the cumulative spectral energy distribution (SED)
 of local active galactic nuclei (AGN) in the 3--300~keV band and
  compare it with the spectrum of the cosmic X-ray background (CXB) in order to
 test the widely accepted paradigm that the CXB is a superposition of AGN and
 to place constraints on AGN evolution. 
}
{We performed a stacking analysis of the hard X-ray spectra of AGN
  detected in two recent all-sky surveys, performed by the IBIS/ISGRI
  instrument aboard INTEGRAL and by the PCA instrument aboard RXTE, 
  taking into account the space densities of AGN with
  different luminosities and absorption column densities.}
{We derived the collective SED of local AGN in the 3--300~keV energy band. 
  Those AGN with luminosities below $10^{43.5}$~erg~s$^{-1}$
  (17--60~keV) provide the main contribution to the local volume hard
  X-ray emissivity, at least 5 times more than more luminous
  objects. The cumulative spectrum exhibits (although with marginal
  significance) a cutoff at energies above $\sim $100--200~keV and is
  consistent with the CXB spectrum if AGN evolve over cosmic time in
  such a way that the SED of their collective high-energy emission has
  a constant shape and the relative fraction of obscured AGN remains
  nearly constant, while the AGN luminosity density undergoes strong
  evolution between $z\sim 1$ and $z=0$, a scenario broadly consistent
  with results from recent deep X-ray surveys.
}
{The first direct comparison between the collective hard X-ray SED of local 
  AGN and the CXB spectrum demonstrates that the popular concept of the CXB 
  being a superposition of AGN is generally correct. By repeating this test
  using improved AGN statistics from current and future hard X-ray
  surveys, it should be possible to tighten the constraints on
  the cosmic history of black hole growth.
}

\maketitle

\section{Introduction}
\label{s:intro}

It is widely believed that the bulk of the cosmic X-ray background (CXB) 
consists of emission from all active galactic nuclei (AGN) in
the visible Universe \citep{setwol89}. Therefore, the CXB represents a
unique integral record of the history of growth of massive black holes
since the early epochs until the present time. The popularity of this
concept rests on the absence of significant distortions in the thermal
spectrum of the cosmic microwave background, which rules out that a
hot intergalactic medium is a major contributor to the CXB
\citep{wrietal94}, and also on the fact that most of the background emission at
energies below several keV has been directly resolved into individual
AGN (see \citealt{brahas05} for a recent review). 

However, the peak of the CXB spectral energy distribution (SED) is situated 
near 30 keV (e.g. \citealt{gruetal99}), where $\sim 99$\% of the background
emission remains unresolved \citep{krietal07}. For this reason,
currently popular CXB synthesis models (e.g. \citealt{giletal07})
compare versatile AGN statistics (source counts, luminosity functions
at different redshifts, distribution of absorption column densities, 
etc.) collected by X-ray surveys at energies below $\sim 8$~keV with the CXB 
spectrum, {\sl under certain assumptions about AGN SEDs in the hard
X-ray and soft gamma-ray bands} (from 10 to several hundred keV). This allows
one to test the overall paradigm of the CXB being a superposition of AGN and 
possibly to get an idea about still missing (e.g. heavily obscured) 
populations of AGN and other types of extragalactic objects. 
 
Recent observations have provided a number of constraints for this picture. In 
particular, it has been established how the
AGN luminosity function defined at rest-frame energies below 8 keV
evolves with redshift, at least out to $z\sim 2$, and there are now
constraints on the fraction of obscured AGN as a function of luminosity and
redshift (e.g. \citealt{uedetal03,baretal05,tozetal06}). Most of these
data were obtained in deep, pencil-beam extragalactic
surveys conducted by the Chandra and XMM-Newton observatories.

Valuable complementary statistics have recently been provided by
shallow hard X-ray surveys of the whole sky performed by RXTE
\citep{revetal04,sazrev04}, Swift \citep{maretal05}, INTEGRAL
\citep{basetal06,becetal06,krietal07,sazetal07} and HEAO-1
\citep{shietal06}. Most importantly, these surveys have
made it possible for the first time to reliably measure the distribution of 
AGN absorption column densities well into the Compton thick regime
($\nh\la 10^{25}$~cm$^{-2}$), albeit only at low redshifts ($z\la
0.1$). In particular, it was found that the observed ratio of obscured
($\nh>10^{22}$~cm$^{-2}$) to unobscured ($\nh<10^{22}$~cm$^{-2}$) AGN
drops from about 2:1 at hard X-ray (17--60~keV) luminosities 
$\la 10^{43.5}$~erg~s$^{-1}$ to about 1:3 at higher luminosities
(\citealt{sazetal07}; see also \citealt{sazrev04,maretal05,shietal06}).

As was mentioned above, previous CXB studies used some  
fiducial intrinsic AGN SED going up to several hundred keV, which
was usually assumed to be similar to the few measured spectra of
brightest nearby Seyfert galaxies (e.g. \citealt{jouetal92,zdzetal95}). 
However, {\sl it has never been demonstrated that the cumulative hard X-ray 
SED of all AGN residing in a given volume of the Universe is 
compatible with the CXB spectrum}. It is only now that such crucial
comparison can be made for the first time, albeit only for the
low-redshift AGN population, using the all-sky hard X-ray surveys
mentioned above. The IBIS/ISGRI instrument \citep{ubeetal03} aboard
INTEGRAL \citep{winetal03} is particularly suitable for this purpose
since it is effectively sensitive up to 300~keV, i.e. well into
the energy range where the collective AGN SED is expected to have
a cutoff if our understanding of the CXB origin is correct.

The purpose of the present work is to estimate the cumulative
hard X-ray SED of local ($z\la 0.1$) AGN. For this
purpose we stack the spectra of AGN detected
in two recent all-sky surveys, performed by the IBIS/ISGRI instrument
aboard INTEGRAL and by the PCA instrument aboard RXTE, taking into
account the space densities of AGN with different luminosities 
(\S\ref{s:stack}). By comparing the derived cumulative SED of local AGN,
which spans two decades in energy (3--300~keV), with the CXB spectrum we 
obtain constraints on the evolution of AGN over cosmic time
(\S\ref{s:cxb}). A cosmology with $\Omega_{\rm m}=0.3$, 
$\Omega_\Lambda=0.7$, and $H_0=75$~km~s$^{-1}$~Mpc$^{-1}$ is adopted
throughout the paper. All quoted uncertainties are 1$\sigma$ unless
noted otherwise. 

\section{Analysis}
\label{s:stack}

We recently used the (mostly) serendipitous all-sky survey conducted
by the IBIS/ISGRI instrument \citep{krietal07} to obtain a sample of
nearby AGN detected in the 17--60~keV energy band
\citep{sazetal07}. This sample is well suited for estimating the cumulative
hard X-ray SED of local AGN. In what follows we first
describe our stacking spectral analysis performed on this AGN set
(\S\ref{s:integral}). We then report on our similar analysis
carried out at lower energies (3--20~keV) using the RXTE Slew Survey
(\S\ref{s:rxte}). Finally we put both sets of results together to obtain a
composite AGN SED covering the energy range 3--300~keV
(\S\ref{s:model}, \S\ref{s:final}).
 
\subsection{INTEGRAL sample}
\label{s:integral}

The all-sky survey reported by \cite{krietal07} is based on
INTEGRAL/IBIS/ISGRI observations performed during 2002--2006, including a
special series of observations of ``empty'' extragalactic fields. It
greatly improves on previous hard X-ray surveys in terms of angular
resolution ($\sim 12^\prime$) and sensitivity. For 80\% of the sky a
source detection limit of $5~{\rm mCrab}\approx 7\times
10^{-11}$~erg~s$^{-1}$~cm$^{-2}$ (17--60~keV) or better is achieved. A
total of 403 sources were found, with at most 1--2 of them being spurious.

Nearly two thirds of the sources in the \cite{krietal07} catalog 
reside in the Galaxy, while the rest are confirmed or suspected AGN. We
previously used the INTEGRAL catalog to construct the hard X-ray (17--60~keV)
luminosity function of nearby AGN and to study their
distribution in absorption column density \citep{sazetal07}. As in
that work, we now use the subsample of AGN located outside
the Galactic plane region ($|b|>5^\circ$) to avoid problems with
identification incompleteness at low Galactic latitudes; our AGN
sample is highly complete ($\sim 95$\%) at $|b|>5^\circ$. 

In fact the sample of AGN used in the present work is
slightly different from that presented by \cite{sazetal07} but 
precisely corresponds to the catalog of \cite{krietal07},
which included a few additional INTEGRAL observations and follow-up
identifications that had become available between those two
publications. Specifically, our current sample is composed of 76,
rather than 74 AGN detected on the IBIS/ISGRI average sky map at
$|b|>5^\circ$. Information on the two additional sources is
presented in Table~\ref{tab:int_agn}, which should be regarded as a
continuation of Table~1 in \cite{sazetal07}. Among these 76 AGN, there
are 8 blazars and 68 nearby ($z\la 0.1$) Seyfert galaxies
(although a few of them still lack an exact optical
classification). Below we focus on these non-blazar AGN, although a
short notice is made in \S\ref{s:final} with respect to the blazar
contribution to the cumulative AGN spectrum.

\begin{table*}
\caption{AGN detected on the average IBIS/ISGRI map at $|b|>5^\circ$
in addition to the \cite{sazetal07} sample
\label{tab:int_agn}
}
\scriptsize

\begin{center}

\begin{tabular}{lcccccccc}
\hline
\hline
\multicolumn{1}{c}{Object} &
\multicolumn{1}{c}{Class$^{\rm a}$} &
\multicolumn{1}{c}{Ref.$^{\rm b}$} &
\multicolumn{1}{c}{$z$} &
\multicolumn{1}{c}{$D$$^{\rm c}$} &
\multicolumn{1}{c}{$F_{\rm 17-60}$} & 
\multicolumn{1}{c}{$\log L_{\rm 17-60}$$^{\rm d}$} &
\multicolumn{1}{c}{$N_{\rm H}$} & 
\multicolumn{1}{c}{Ref.$^{\rm e}$} \\
   
\multicolumn{1}{c}{} &     
\multicolumn{1}{c}{} &     
\multicolumn{1}{c}{} &     
\multicolumn{1}{c}{} &
\multicolumn{1}{c}{Mpc} & 
\multicolumn{1}{c}{10$^{-11}$~erg~s$^{-1}$~cm$^{-2}$} &       
\multicolumn{1}{c}{erg~s$^{-1}$} &     
\multicolumn{1}{c}{10$^{22}$~cm$^{-2}$} &
\multicolumn{1}{c}{} \\
\hline
SWIFT J0601.9$-$8636=ESO 005-G004 & S2 & 1 & 0.0062 & 22.4 &
$2.51\pm0.46$ & 42.18 & $\sim 100$& 1 \\
IGR J14561$-$3738=ESO 386-G034    & S2 & 2 & 0.0246 &      &
$1.40\pm0.26$ & 43.23 & $>100$ & 2\\  
\hline
\end{tabular}
\end{center}

$^{\rm a}$ Optical AGN class: S2 -- Seyfert 2 galaxy.

$^{\rm b}$ Reference for the optical classification: (1)
\cite{moretal06}, (2) \cite{masetal07}. 

$^{\rm c}$ Distance according to the Nearby Galaxies Catalogue
\citep{tully88}.

$^{\rm d}$ Observed (uncorrected for absorption) luminosities in the
17--60~keV energy band.

$^{\rm e}$ Quoted absorption column density $\nh$ is adopted from: (1)
\cite{uedetal07}, (2) \cite{sazetal08}.

\end{table*}

\subsubsection{Reconstruction of multiband sky images}

To obtain the hard X-ray spectra of the INTEGRAL AGN 
we essentially repeated our analysis of IBIS/ISGRI data, previously
performed in the 17--60~keV energy band (see \citealt{krietal07} for
details), in 7 narrow channels: 17--26, 26--38, 38--57, 57--86, 86--129,
129--194 and 194--290 keV. This resulted in time-averaged maps
of the whole sky in these energy bands, which were then used to
measure source spectral fluxes. In this analysis, the source positions 
were fixed at the values determined by \cite{krietal07} in the 17--60~keV 
band.

There are two important differences with respect to the
\cite{krietal07} study. First, for the present work we updated our
data set by adding all IBIS/ISGRI data that became available to us
since the last publication. This amounted to 26~Ms of cleaned
and deadtime-corrected data in addition to 33~Ms available before.
Secondly, in reconstructing clean sky images from raw 
IBIS/ISGRI shadowgrams in a given energy channel we used only those sources 
detectable in that band rather than the entire source catalog. This
essentially eliminates iterative removal of sources from images,
unnecessary in the high-energy channels (129--194 and 194--290 keV),
where only 27 and 8 sources, respectively, are detected ($>5\sigma$) on
the whole sky, and reduces the error in measured fluxes. 

It is important to emphasize that despite the addition of the most recent
IBIS/ISGRI data we continued to base our analysis on the original AGN
catalog (\citealt{sazetal07} plus the two sources in
Table~\ref{tab:int_agn}) and the corresponding exposure map from 
\cite{krietal07}. Thus, the new data, added in an attempt to improve the 
quality of the source spectra, do not affect the statistical properties of 
our AGN sample.

When carrying out a stacking analysis like the one reported here 
one should always worry that some systematic uncertainties associated
with the reconstruction of individual source fluxes may be greatly
enhanced by adding up the fluxes of many sources. Therefore, to verify
that our results do not suffer from systematic uncertainties related
to image reconstruction we made a number of simulations.

In particular, we have extracted spectra from 3,000 ``empty''
positions on the sky chosen in accordance with the INTEGRAL
exposure map, i.e. the number density of these ``zero-flux'' sources
was made higher in regions of large exposure than in regions of small
exposure. When we carried out simulations including in the iterative
source removal procedure all the sources from the INTEGRAL catalog,
irrespective of the significance of their detection in the considered 
energy channel, the mean of the distribution of the fluxes of the simulated
sources proved to be $\sim0.05$--0.1$\sigma$ above zero 
(where $\sigma$ is the typical uncertainty of source flux measurement), 
which is unacceptable for our stacking analysis. This
result is actually anticipated for the employed algorithm of image
reconstruction \citep{krietal07}. However, when we repeated the simulations in
exactly the same manner as we carried out our real stacking analysis,
i.e. including into our iterative source removal procedure only those
INTEGRAL sources that are significantly detected in the considered energy
channel, the distribution of the fluxes of the simulated
sources proved to be consistent with the normal one with zero mean and the 
dispersion expected from Poisson statistics. 

This proves that a stacking analysis of as many as 3,000 sources does not
 suffer from any systematic
uncertainties in our image reconstruction procedure. 

\subsubsection{Stacked spectra}

The individual AGN spectra derived from the reconstructed IBIS/ISGRI multiband
sky images will be discussed in detail elsewhere. Here we are only
interested in the average properties of these spectra. 

\begin{figure}
\centering
\includegraphics[width=\columnwidth]{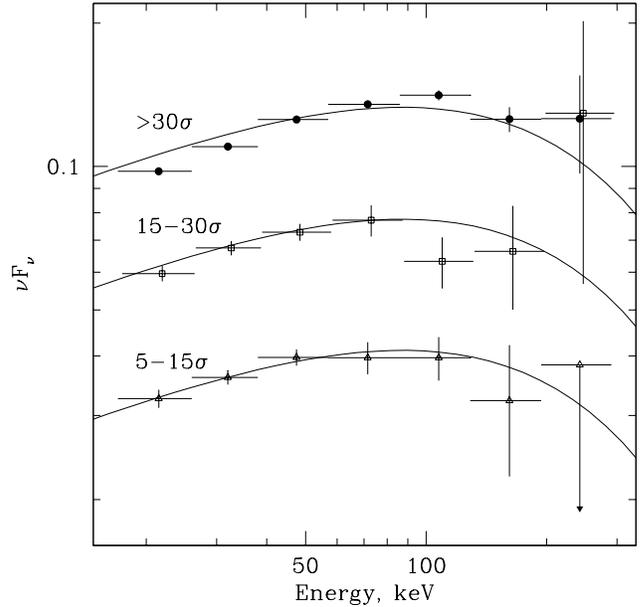}
\caption{Average IBIS/ISGRI spectra of non-blazar AGN divided into
three groups according to their detection significance in the
17--60~keV energy band. The spectral normalizations are
arbitrary. Upper limits are $1\sigma$. The solid lines show crude
approximations by the cutoff power law model with $\Gamma= 1.65$ and
$E_f=250$~keV.
}
\label{fig:threespectra}
\end{figure}

We first separately stacked AGN spectra within three groups defined by
the source signal-to-noise ratio in the 17--60~keV energy band: $>30$ (5
sources), between 15 and 30 (6 sources) and $<15$ (57 sources). This
is done mainly to prevent the stacked spectrum from being dominated by
one or two bright sources. The resulting spectra are shown in
Fig.~\ref{fig:threespectra}. One can see that all three spectra are
similar to each other. They exhibit some rollover between $\sim 40$
and $\sim 200$~keV and can be approximately described by a power law
with a high-energy cutoff: $dN/dE \propto 
E^{-\Gamma}\exp{(-E/E_f)}$, with $\Gamma\sim 1.65$ and $E_f\sim
250$~keV (see the solid lines in Fig.~\ref{fig:threespectra}). This
model provides an adequate approximation of the observed spectra, but
given the limited energy coverage (17--290~keV), other functional
forms (e.g. a broken power law) would fit the INTEGRAL data equally
well.

The fact that the averaged spectra of local AGN presented in
Fig.~\ref{fig:threespectra} have a maximum at energies $\sim
80$--100~keV, which can be explained by the presence of a high-energy
cutoff with $E_f\sim 250$~keV, implies that if the same hard X-ray
emission were coming from redshifts $z\sim 1$--2 its spectrum observed at 
$z=0$ would have a peak near $\sim 30$~keV, similar to the cosmic X-ray 
background. However, these stacked spectra cannot yet be regarded as
characteristic of the collective hard X-ray emission of the local AGN
population, since neither AGN space densities nor luminosities were
taken into account when constructing them. 

To really obtain the cumulative SED of local AGN, the
following weighted stacking should be done: 
\beq
S_i=\sum_{j}\frac{L_{i,j}}{V_{{\rm max},j}}=
\sum_{j}\frac{4\pi D_j^2 F_{i,j}}{V_{{\rm max},j}}.
\label{eq:si}
\eeq
Here $S_i$ is the cumulative volume emissivity (measured in units of
erg~s$^{-1}$~Mpc$^{-3}$) in energy channel $i$ (from 1 to 7),
$L_{i,j}$ is the observed luminosity of $j$'th AGN in channel
$i$, $F_{i,j}$ is the measured flux from $j$'th AGN in channel $i$,
$D_j$ is the source luminosity distance and $V_{{\rm max},j}$ is the
volume of space within which an AGN with observed luminosity $L_{{\rm hx},j}$ 
in the 17--60~keV band could be detected in the INTEGRAL survey. 
We adopted the source distances ($D_j$) and signal-to-noise ratios
(used to derive the maximum volumes $V_{{\rm max},j}$ given the
INTEGRAL sky exposure map) from \cite{sazetal07}, where the hard X-ray
luminosity function of local AGN was derived in a similar way, and also from
Table~\ref{tab:int_agn}. The conversion of detector counts to photon
fluxes in the 7 spectral bands was calibrated using IBIS/ISGRI
measurements of the Crab and assuming that the Crab spectrum is given
by $dN/dE=10(E/1~{\rm keV})^{-2.1}$~phot~keV$^{-1}$~s$^{-1}$~cm$^{-2}$. 
In calculating the $V_{{\rm max},j}$ volumes we also assumed the observed
spectra to be Crab-like. Although in general this may be a poor assumption 
for heavily obscured ($\log\nh\gg 10^{24}$~cm$^{-2}$) AGN, it is unlikely to 
introduce a significant error in our case, because we use a relatively 
narrow energy band (17--60~keV) for source detection and 
since such strongly absorbed AGN provide a small contribution to the 
resulting cumulative SED.

There are two types of uncertainty associated with the summed spectrum
$S_i$. One results from the errors $\delta F_{i,j}$ in the measured
source spectral fluxes (and the associated uncertainties, $\delta
L_{i,j}=4\pi D_j^2\delta F_{i,j}$, in the spectral luminosities) and
is given by:
\beq
\delta S_{i,1}=\sqrt{\sum_{j}\left(\frac{\delta L_{i,j}}{V_{{\rm
      max},j}}\right)^2}.  
\label{eq:ds1}
\eeq
Another uncertainty is associated with the finite size of our AGN
sample:  
\beq
\delta S_{i,2}=\sqrt{\sum_{j}\left(\frac{L_{i,j}}{V_{{\rm max},j}}\right)^2}. 
\label{eq:ds2}
\eeq
Therefore, these uncertainties are associated with the Poisson distribution of
detector counts and the number of sources, respectively. Since the first 
uncertainty is directly linked to the survey's sensitivity in
different energy bands, its relative amplitude is much larger for the
high-energy channels than for the low-energy ones. In contrast, the
second uncertainty is nearly independent of the energy band ($\delta
S_{i,2}/S_i\approx{\rm const}$) and affects the normalization of the
cumulative SED rather than its shape. 

\begin{figure}
\centering
\includegraphics[width=\columnwidth]{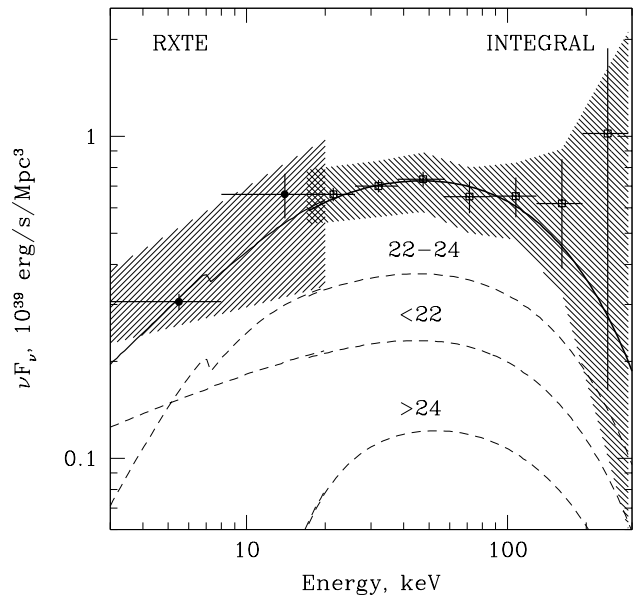}
\caption{Cumulative 3--300~keV SED of local AGN with luminosities
(observed in the 17--60 keV band or absorption corrected in the
  3--20~keV band) $10^{41}<\lhx<10^{43.5}$~erg~s$^{-1}$, obtained with 
INTEGRAL/IBIS/ISGRI and RXTE/PCA. The error bars are 1$\sigma$
uncertainties of the first type [equation~(\ref{eq:ds1})],
while the shaded regions show the combined 1$\sigma$ uncertainties of
the first and second type [equation~(\ref{eq:ds2})]. The solid line is
the best-fit model by a sum of absorbed and unabsorbed power laws with
a high-energy exponential cutoff [equation~(\ref{eq:model}),
  Table~\ref{tab:bestfit}]. The dashed lines indicate the contributions
of unobscured AGN, and obscured AGN with $\nh<10^{24}$ and
$>10^{24}$~cm$^{-2}$. The RXTE points have been mutliplied by a factor
of 1.1 and the INTEGRAL points divided by the same factor to correct
for the effect of the local large-scale structure (see \S\ref{s:lss}).
}
\label{fig:bestfit_l41_43.5}
\end{figure}

\begin{figure}
\centering
\includegraphics[width=\columnwidth]{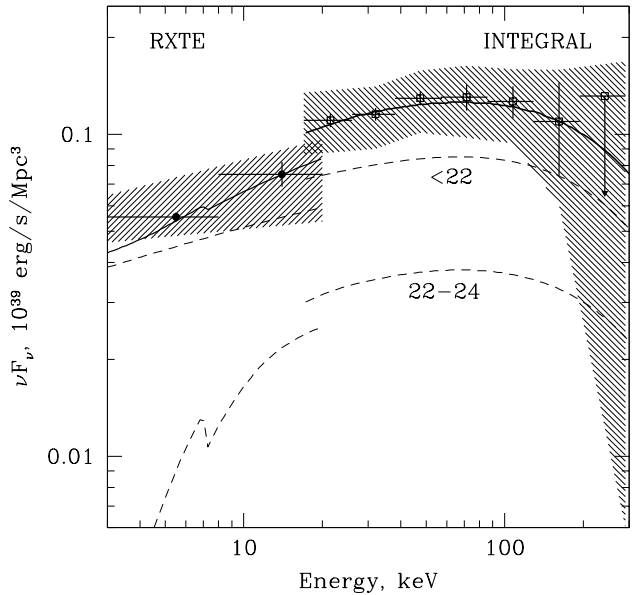}
\caption{Cumulative 3--300~keV SED of local AGN with
$\lhx>10^{43.5}$~erg~s$^{-1}$; see Fig.~\ref{fig:bestfit_l41_43.5} for
explanations. The upper limit in the highest energy channel is
1$\sigma$. We note the apparent mismatch between the INTEGRAL and RXTE
parts of the SED, which is statistically insignificant and may be
attributable in part to effects of the local large-scale structure,
which we cannot accurately estimate and thus have not corrected for in
this high-luminosity case (see in \S\ref{s:lss} and \S\ref{s:model}).
}
\label{fig:bestfit_lg43.5}
\end{figure}

In Figs.~\ref{fig:bestfit_l41_43.5} and \ref{fig:bestfit_lg43.5}
we show two cumulative AGN SEDs obtained using
equations~(\ref{eq:si})--({\ref{eq:ds2}). In the first case the 
stacking was done for the AGN (40 in total) with luminosities between
$\lmin=10^{41}$ and $\lmax=10^{43.5}$~erg~s$^{-1}$ 
(17--60~keV). In the other case, $\lmin=10^{43.5}$~erg~s$^{-1}$
and $\lmax=\infty$ (27 AGN). This division approximately corresponds
to the low-luminosity and high-luminosity branches of the luminosity
function of nearby AGN, which are characterized by different ratios of
obscured and unobscured sources \citep{sazetal07} and so may represent
somewhat different populations of AGN. Note that one source in
our sample -- the Seyfert galaxy NGC~4395 -- has luminosity less
than $10^{41}$~erg~s$^{-1}$ ($\lhx\sim 40.4$) and was not used in the
stacking analysis.

\subsection{RXTE/PCA sample}
\label{s:rxte}

We have thus obtained the cumulative SEDs of
local low-luminosity and high-luminosity AGN in the energy range
17--290~keV. There is an obvious need to continue these spectra to 
lower energies where intrinsic X-ray absorption in AGN should come
into play, producing a low-energy cutoff in the spectra of obscured
objects. There are two possible ways to obtain this low-energy part of the
cumulative SED. One is to search the literature and data archives
for X-ray spectra of the INTEGRAL AGN. We decided not to follow
this approach for two reasons. First, published or archival X-ray
observations of sufficient quality do not exist for all of our
objects. Second, because AGN are inherently variable, adding
lower-energy spectral fluxes to the IBIS/ISGRI hard X-ray spectra
could introduce a bias due to the fact that these particular AGN tended to be
detected by IBIS/ISGRI as they were bright relative to their typical
flux levels (averaged over many years), while a different X-ray
telescope will find the same source equally likely either below or
above its average flux level (see \citealt{shietal06} for a detailed
discussion of this effect}). As a result, the low-energy part of the
cumulative SED derived by this method would somewhat
underestimate the true SED of the local AGN
population. 

Another feasible approach consists of using an independent survey of nearby AGN
performed at the energies of our current interest, i.e. just below
20~keV. Such a survey of the whole sky in the energy band 3--20~keV
was recently performed with the PCA instrument aboard the RXTE
observatory during its slews between pointed observations -- the RXTE
Slew Survey, or XSS \citep{revetal04}. In this survey a flux limit of
$1.8\times 10^{-11}$~erg~s$^{-1}$~cm$^{-2}$ (3--20~keV) or better was
achieved for 80\% of the sky at $|b|>10^\circ$. Therefore, the XSS is
substantially more sensitive than the IBIS/ISGRI survey with respect
to unobscured AGN (e.g. Seyfert 1 galaxies) but is less sensitive with
respect to strongly obscured ($\nh\ga 10^{23}$~cm$^{-2}$) ones.

\begin{table*}
\caption{Non-blazar AGN detected at $|b|>10^\circ$ during the RXTE Slew Survey in
  addition to the \cite{sazrev04} sample}
\label{tab:rxte_agn}

\scriptsize

\begin{center}

\begin{tabular}{lllccccccrc}
\hline
\hline
\multicolumn{1}{c}{XSS object} &
\multicolumn{1}{c}{Name} &
\multicolumn{1}{c}{Class$^{\rm a}$} & 
\multicolumn{1}{c}{Ref.$^{\rm b}$} & 
\multicolumn{1}{c}{3--8 keV} & 
\multicolumn{1}{c}{8--20 keV} & 
\multicolumn{1}{c}{$z$} &
\multicolumn{2}{c}{$\log L_{3-20}$$^{\rm c}$} &
\multicolumn{1}{c}{$N_{\rm H}$} & 
\multicolumn{1}{c}{Ref$^{\rm d}$} \\
   
\multicolumn{1}{c}{(J2000.0)} &
\multicolumn{1}{c}{} &     
\multicolumn{1}{c}{} & 
\multicolumn{1}{c}{} & 
\multicolumn{1}{c}{cnt s$^{-1}$} &       
\multicolumn{1}{c}{cnt s$^{-1}$} &  
\multicolumn{1}{c}{} &
\multicolumn{2}{c}{erg s$^{-1}$} &     
\multicolumn{1}{c}{10$^{22}$ cm$^{-2}$} &
\multicolumn{1}{c}{} \\          
\hline
05054$-$2348 &2MASX J05054575$-$2351139 &S2&1& $0.69\pm0.19$ &
$0.92\pm0.23$ & 0.0350 & 43.77 & 43.83 & 6 & 1 \\ 
12303$-$4232 &IRAS F12295$-$4201        &S1.5&2& $0.48\pm0.09$ &
$0.29\pm0.11$ & 0.1000 & 44.37 & 44.37 & $<1$ & 1 \\    
12389$-$1614 &2MASX J12390630$-$1610472 &S2&3& $0.93\pm0.11$ &
$0.51\pm0.13$ & 0.0367 & 43.75 & 43.77 & 2 & 2 \\ 
15076$-$4257 &2MASX J15080462$-$4244452 &S1&4& $0.73\pm0.05$ &
$0.24\pm0.06$ & 0.0565 & 43.96 & 43.96 & $<1$ & 3 \\  
18236$-$5616 &IC 4709                   &S2&5& $0.32\pm0.09$ &
$0.51\pm0.12$ & 0.0169 & 42.86 & 42.98 & 12 & 1 \\   
19459+4508   &2MASX J19471938+4449425   &S2&1& $0.38\pm0.10$ &
$0.38\pm0.12$ & 0.0532 & 43.84 & 43.95 & 11 & 2 \\ 
21354$-$2720 &IRAS F21318$-$2739      &S1.5&1& $0.41\pm0.11$ &
$0.32\pm0.13$ & 0.0670 & 43.99 & 43.99 & $<1$ & 4 \\    
\hline
\end{tabular}

\end{center}
 
$^{\rm a}$ Optical AGN class: S1.5 -- Seyfert 1.5 galaxy, S2 --
Seyfert 2 galaxy.

$^{\rm b}$ Reference for the optical classification: (1)
\cite{biketal06}, (2) \cite{lanetal07}, (3) \cite{masetal06a}, (4) a
Seyfert 1 nucleus is suggested by the 6dF spectrum, (5) \cite{masetal06b}.

$^{\rm c}$ Left and right columns give the observed and absorption
corrected luminosities in the 3--20~keV band, respectively.

$^{\rm d}$ Quoted absorption column density value is adopted from or
based on: (1) \cite{revetal06},(2) \cite{sazetal05}, (3) ROSAT
data, (4) Swift data.

\end{table*}

The high Galactic latitude ($|b|>10^\circ$) part of the XSS was
previously used to obtain a catalog of AGN, to construct the 
3--20~keV luminosity function of nearby AGN and to study their
distribution of absorption column densities \citep{sazrev04}, quite
similarly to the analysis of the INTEGRAL survey we carried out
later. Here we use the XSS to obtain low-energy intensities for our
cumulative hard X-ray SEDs of local AGN.

Since the original XSS publications several
changes have taken place that need to be taken into account
here. First, originally a substantial fraction ($\sim 30$\%) of XSS
sources were unidentified, mainly because of their poor 
localization (uncertainty up to $1^\circ$). Over the passed three 
years many of these sources have been identified, thanks to dedicated
follow-up efforts \citep{biketal06,revetal06,lanetal07} or due
to detection and improved localization of some of these sources by INTEGRAL 
and/or Swift. As a result, the XSS sample of AGN detected ($>4\sigma$)
at $|b|>10^\circ$ is now composed of 103, rather than 95 sources,
including 84 non-blazar AGN. The new non-blazar AGN are listed in
Table~\ref{tab:rxte_agn}, which should be considered a continuation
of Table~1 in \cite{sazrev04}. At the same time the number of
unidentified XSS sources (detected at $>4\sigma$) has decreased to 16
objects. Therefore, the XSS AGN sample is now at least 82\%
complete\footnote{Here it is also taken into account that $\sim 6$
additional AGN are possibly confused with other XSS sources.}, so
incompleteness is significantly less of a problem than before. 

Secondly, through careful testing of our methods developed for
analyzing data recorded in the slew mode of RXTE observations we 
found that the conversion factors from PCA counts to source photons quoted in
\cite{revetal04} had been underestimated by a factor of $\sim 1/0.7$,
because we had not taken into account that the PCA field of view
is rapidly moving across sources during RXTE slews and used a slow movement
approximation of the PCA response. This implies that, to a first
approximation, the AGN luminosities quoted in \cite{sazrev04} should
be revised upwards by a factor of 1/0.7 (the same correction factor applies to
Table~\ref{tab:rxte_agn} presented here). 

There are a few other minor modifications that have been implemented,
including the correction of a misprint in \cite{sazrev04} for the
redshift of NGC~7582 from 0.053 to 0.0053 and accordingly revising its
luminosity. 

We plan to release an updated version of the XSS catalog in the near future, 
including the modifications described above and a number of additional source 
identifications. 

It is important to note that although the RXTE and INTEGRAL AGN
samples are comparable in size, they overlap by just $\sim 30$\%,
i.e. these samples are mostly independent of each other. Although this
result may seem unexpected, it can readily be explained by a
combination of circumstances: 1) the XSS catalog is defined at
$|b|>10^\circ$, whereas the INTEGRAL AGN sample used here is defined
at $|b|>5^\circ$ and the strip of the sky between $|b|=5^\circ$ and
$|b|=10^\circ$ has been well covered by INTEGRAL observations, 2) the
distribution of exposure over the sky is significantly different for the two
surveys, 3) due to its lower energy band the 
XSS is biased toward detecting unobscured and weakly obscured AGN and
against detecting Compton thick AGN, 4) AGN are inherently variable
and the two surveys were conducted at significantly different epochs. 

We used the updated XSS AGN sample to obtain the 3--8 and 8--20~keV
data points for our cumulative hard X-ray SEDs of local low-luminosity
and high-luminosity AGN (Figs.~\ref{fig:bestfit_l41_43.5} and
\ref{fig:bestfit_lg43.5}) following the same $1/V_{\rm max}$ weighted
summing method as we used above (\S\ref{s:integral}) to obtain the
high-energy parts of the SEDs based on IBIS/ISGRI data. We note
that the XSS sample includes 6 non-blazar AGN located at $z\sim
0.15$--0.3. Nonetheless, since the vast majority of the sample are
truly local sources ($z\la 0.1$) and the several more distant AGN
contribute just a few per cent to the resulting cumulative SEDs, we
included these objects in our stacking analysis as if they were local,
i.e. ignoring any evolution of AGN between $z=0.3$ and $z=0$.

There are three differences with respect to our analysis of the IBIS/ISGRI
data, all caused by the strong effect of intrinsic absorption on
AGN spectra below 20~keV. First, in determining the $V_{\rm max}$ values
for the XSS sources, we took into account their absorption column
densities given in Table~1 in \cite{sazrev04} and in
Table~\ref{tab:rxte_agn} presented here. That is $V_{\rm  max}$ is now
defined as the volume of space in which the XSS would detect an AGN
with its observed 3--20~keV luminosity and column density $\nh$ (not
just the luminosity as in the INTEGRAL case). Second, the $\nh$
values were taken into account also to calculate the observed and
intrinsic (corrected for the absorption) luminosities of the XSS AGN
in the subbands 3--8 and 8--20~keV from the detector counts measured 
in these channels (as given in Table~1 in \cite{sazrev04} and in
Table~\ref{tab:rxte_agn} presented here). In doing this conversion the source
spectrum was assumed to have an intrinsic slope of 1.8, although the
result only weakly depends on this assumption \citep{revetal04}.
And finally, we used the same two luminosity ranges: 1) $\lmin=10^{41}$,
$\lmax=10^{43.5}$~erg~s$^{-1}$ (38 AGN) and 2) $\lmin=10^{43.5}$,
$\lmax=\infty$ (46 AGN), as for our stacking analysis of INTEGRAL
spectra, but for RXTE these ranges are defined for intrinsic (i.e. 
absorption-corrected) luminosities 
in the 3--20~keV energy band. This ensures that the stacking of both
INTEGRAL and RXTE spectra is done for almost identical populations of
AGN, since the unabsorbed luminosities of AGN in the 3--20~keV and 17--60~keV
bands are very similar (see e.g. Fig.~7 in \citealt{sazetal07}).

\subsubsection{Influence of the local large-scale structure}
\label{s:lss}

Both the INTEGRAL and RXTE all-sky surveys are characterized by
substantially nonuniform coverage of the sky. Furthermore, the sky
exposure maps of these surveys are quite different. Because both
surveys effectively picked out AGN within $\sim$ 
100--200~Mpc from us and the distribution of matter in the local
Universe is inhomogeneous, effects of the local large-scale structure (LSS)
may come into play when comparing AGN space densities or any
quantities depending thereof (luminosity functions, cumulative SEDs etc.) 
estimated using one survey and the
other. It is straightforward to estimate these effects if we
assume that AGN are distributed in the local Universe approximately as
normal galaxies. \cite{krietal07} have actually demonstrated that the
AGN detected by INTEGRAL do follow quite well the spatial distribution
of galaxies. 

To assess the effect of the local LSS on the cumulative SEDs of low-
and high-luminosity AGN
(Figs.~\ref{fig:bestfit_l41_43.5} and \ref{fig:bestfit_lg43.5}), we made
use of the IRAS PSCz catalog \citep{sauetal00}, which is the most complete
published all-sky galaxy redshift catalog. Our analysis essentially
consisted of counting PSCz galaxies within the volume of space in
which the INTEGRAL or RXTE survey can detect AGN with a given hard 
X-ray (3--20 or 17--60~keV) luminosity, and calculating 
the average galaxy number density in that volume of the Universe.
To avoid problems with incompleteness of the PSCz catalog at low Galactic
latitudes \citep{sauetal00}, we excluded the sky below $|b|=10^\circ$
from the analysis, even though our INTEGRAL AGN sample extends down to 
$|b|=5^\circ$. In reality, the situation is somewhat more complicated,
first because the INTEGRAL and RXTE surveys are nonuniform, hence the
distance out to which a source with a given luminosity is detectable
depends on the direction in the sky, and also because the PSCz catalog
itself becomes incomplete at a certain distance that depends on galaxy infrared
luminosity. Nonetheless, it is not difficult to take these
effects into account to first approximation. 

Specifically, we first constructed 9 subsamples of IRAS PSCz galaxies
according to their infrared (60~$\mu$m) luminosity: $L_{\rm
  IR}>L_{{\rm cut},i}=5\times 10^8$, $10^9$, 
$2\times 10^9$, $5\times 10^9$, $10^{10}$, $2\times 10^{10}$, $5\times
10^{10}$, $10^{11}$, $2\times 10^{11} L_\odot$. These subsamples
are expected to be complete out to $D_{\rm max}=23$, 32,
45, 72, 102, 144, 228, 322, 455~Mpc, since the PSCz catalog is
highly complete at 60~$\mu$m fluxes higher than 0.6~Jy
\citep{sauetal00}, except near the Galactic plane (we ignore here any
$k$-corrections since we are dealing with low redshifts). We then
calculated the average number density of IRAS PSCz galaxies with
$L_{\rm IR}>L_{{\rm cut},i}$ ($i=1$...9) within the volumes of space
$V(\lhx)$ probed by INTEGRAL or RXTE at different hard X-ray luminosities
(17--60 or 3--20~keV, respectively): $\log\lhx=41$, 41.5, 42, 42.5,
43, 43.5, 44.  For simplicity, in calculating $V(\lhx)$ we assumed
power-law spectra with a photon index of 1.8. Of all the resulting 
galaxy density estimates we finally took into account only those satisfying
the condition $D_{\rm max}(L_{{\rm cut},i})\ge D_{\rm 
 90}(\lhx)$, where $D_{\rm 90}$ is the distance within which 90\% of
the space volume probed by INTEGRAL or RXTE at $\lhx$ is contained. 

\begin{figure}
\centering
\includegraphics[width=\columnwidth]{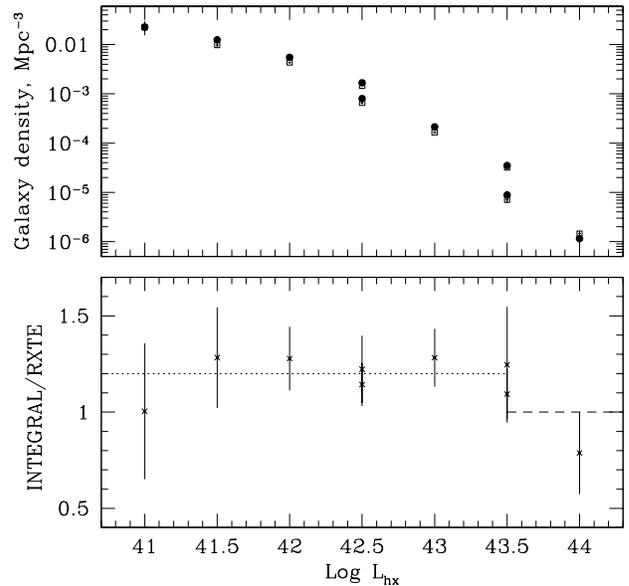}
\caption{{\sl Upper panel:} Average number density of IRAS PSCz
galaxies within the volumes of space probed at $|b|>10^\circ$ by
INTEGRAL (filled points) and RXTE (open points) as a function of AGN
hard X-ray luminosity. Each point is based on one of 9
subsamples of IRAS PSCz galaxies (from top to bottom): $L_{\rm
  IR}>L_{\rm cut}=5\times 10^8$, $10^9$, $2\times 10^9$, $5\times
10^9$, $10^{10}$, $2\times 10^{10}$, $5\times 10^{10}$, $10^{11}$,
$2\times 10^{11} L_\odot$. For a given AGN luminosity only one or
two best (accurate to within $\sim$10--30\%) density estimates are
presented. See text for further explanations. {\sl Lower panel:}
Ratio of the densities in the 
INTEGRAL and RXTE volumes as a function $\lhx$. Also shown are the
typical ratios, 1.2 and 1.0, at $\lhx<10^{43.5}$  (dotted line) and
$\lhx>10^{43.5}$~erg~s$^{-1}$ (dashed line), respectively, which are
adopted as correction factors for the cumulative SEDs of low- and
high-luminosity AGN, respectively.
}
\label{fig:dens_int_rxte}
\end{figure}

We have thus used a combination of dwarf, normal and giant IRAS galaxies
to compare the average galaxy number densities characteristic
of the INTEGRAL and RXTE surveys in the broad range of AGN
luminosities of our interest. The result of this comparison is shown
in Fig.~\ref{fig:dens_int_rxte}. It indicates that the local volume of 
Universe probed by INTEGRAL at $\lhx\la 10^{43.5}$~erg~s$^{-1}$ is a
factor of $\sim 1.2$ denser than the corresponding volume probed by
RXTE. A comparison of the sky exposure maps of INTEGRAL
\citep{krietal07} and RXTE \citep{revetal04} provides 
a likely explanation for this: while INTEGRAL has spent a lot of
time observing such major galaxy concentrations as the
Centaurus cluster--Shapley supercluster and the Perseus--Pisces
supercluster, the RXTE Slew Survey has relatively low exposure in
these regions. On the other hand, we do not find (see
Fig.~\ref{fig:dens_int_rxte}) any significant difference in the 
average galaxy number densities of the INTEGRAL and RXTE survey
volumes at higher AGN luminosities ($\lhx\ga 10^{43.5}$). This is also an
expected result, since high-luminosity AGN are typically observed by 
INTEGRAL and RXTE at distances $\sim 100$--200~Mpc and on
such spatial scales the Universe is believed to be approximately
homogeneous. We note however that the limited statistics of 
distant giant galaxies in the IRAS PSCz catalog does not allow us to
estimate the galaxy number density for the INTEGRAL and RXTE
surveys at high $\lhx$ to an accuracy better than $\sim 20$\%.

To take into account to a first approximation the apparent effect of the local 
LSS and assuming that the INTEGRAL and RXTE surveys equally over- and 
underestimate, respectively, the average density in the local Universe, we 
multiplied the RXTE points of the cumulative SED of low-luminosity AGN
(Fig.~\ref{fig:bestfit_l41_43.5}) by a factor of 1.1 and divided the
INTEGRAL points by the same factor. No such corrections were done 
for the cumulative SED of high-luminosity AGN (Fig.~\ref{fig:bestfit_lg43.5}).

\subsection{Spectral modeling}
\label{s:model}

We next modeled, using XSPEC \citep{arnaud96}, the broad-band
(3--300~keV) cumulative SEDs of local
low- and high-luminosity AGN obtained in the previous sections by a
sum of absorbed and unabsorbed power-law spectra with a high-energy
exponential cutoff:
\beq
E^2dN/dE=A\sum_i k_i f(N_{\rm H,i})E^{-\Gamma+2}\exp(-E/E_f).
\label{eq:model}
\eeq

This model is by definition a sum of identical 
spectra absorbed by different column densities $N_{\rm H,i}$ of
neutral gas: 0, $10^{22.25}$, $10^{22.75}$, $10^{23.25}$,  
$10^{23.75}$, $10^{24.25}$, $10^{24.75}$~cm$^{-2}$. We fix the
weights $k_i$ ($\sum k_i=1$) of these columns at the relative
fractions of AGN with different $\nh$ as measured by INTEGRAL
\citep{sazetal07}, ignoring the associated uncertainties. The
corresponding fractions at low ($\lhx< 10^{43.5}$~erg~s$^{-1}$) and
high ($\lhx> 10^{43.5}$~erg~s$^{-1}$) luminosities are given in
Table~\ref{tab:nh}. We use them accordingly for modeling the cumulative
SEDs of low- and high-luminosity AGN
(Fig.~\ref{fig:bestfit_l41_43.5} and Fig.~\ref{fig:bestfit_lg43.5}). 

\begin{table}
\caption{Observed relative fractions of local AGN with different
absorption column densities at low and high luminosities, updated from the 
values reported by \cite{sazetal07}
\label{tab:nh}
}
\begin{center}

\begin{tabular}{ccc}
\hline
\hline
\multicolumn{1}{c}{$\log\nh<22.0$} &
\multicolumn{1}{c}{$k$ ($41<\log\lhx<43.5$)} &
\multicolumn{1}{c}{$k$ ($\log\lhx> 43.5$)} \\

\hline
$<22.0$    & 0.300 & 0.68\\
22.0--22.5 & 0.175 & 0.08\\
22.5--23.0 & 0.075 & 0.04\\
23.0--23.5 & 0.125 & 0.08\\
23.5--24.0 & 0.150 & 0.12\\
24.0--24.5 & 0.100 & 0.00\\
$>24.5$    & 0.075 & 0.00\\
\hline
\end{tabular}

\end{center}

\end{table}

The photoabsorption modifiers $f(N_{{\rm H},i})$ are determined using 
the XSPEC model phabs assuming solar element abundances. Note that
for substantially Compton thick sources ($\nh\ga
10^{24.5}$~cm$^{-2}$) this simple photoabsorption model
becomes inadequate, but since the relative fraction of such sources according
to the INTEGRAL survey is small ($<10$\%) we apply this model
to all AGN. Also, it would be more appropriate to use in
equation~(\ref{eq:model}) the true, i.e. corrected for selection
effects, fractions of AGN with different absorption column densities
rather than the fractions as observed by INTEGRAL in the 17--60~keV energy
band (Table~\ref{tab:nh}). However, for the adopted photoabsorption model
the observed fraction will significantly (by $>10$\%) underestimate
the intrinsic fraction only at $\nh\ga 10^{24.5}$~cm$^{-2}$ and since
the contribution of such heavily obscured AGN to the total SED is
expected to be small ($<$10\%) compared to the various uncertainties
associated with our analysis and also because the photoabsorption
model itself becomes inadequate in the Compton thick regime, we do not
try to correct our spectral fitting procedure for selection effects.

The free parameters of our model are the power-law slope
$\Gamma$, the position of the high-energy cutoff $E_f$ and the
normalization $A$, which is allowed to take different values for the
RXTE and INTEGRAL parts of the spectrum to prevent some systematic
effects, e.g. associated with the cross-calibration of 
the IBIS/ISGRI and PCA detectors or with the local LSS (see \S\ref{s:lss}), 
from affecting the modeling of the
spectral shape. At the same time this allows us not to take into
account the uncertainties associated with the finite sizes of the INTEGRAL and
RXTE AGN samples [equation~(\ref{eq:ds2})] when fitting the spectrum,
since these uncertainties are strongly correlated between the
different energy channels for a given survey and do not affect the
spectral shape. We do however consider these errors when we discuss
the final spectral normalization below.

\begin{figure}
\centering
\includegraphics[angle=-90,width=\columnwidth]{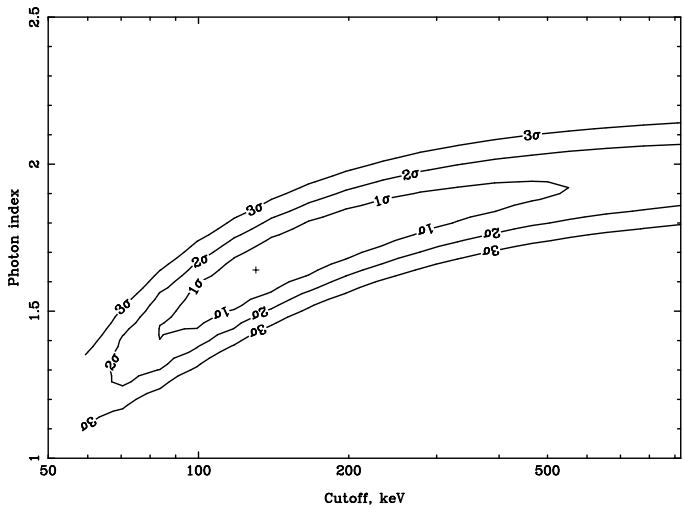}
\caption{Confidence regions for the power-law slope $\Gamma$ and position of
the high-energy cutoff $E_f$ in the model [equation~(\ref{eq:model})] for
the cumulative SED of low-luminosity
($10^{41}<\lhx<10^{43.5}$~erg~s$^{-1}$) AGN. The position of the
best-fit parameter values is indicated by the cross.
} 
\label{fig:contour_l41_43.5}
\end{figure}

\begin{figure}
\centering
\includegraphics[angle=-90,width=\columnwidth]{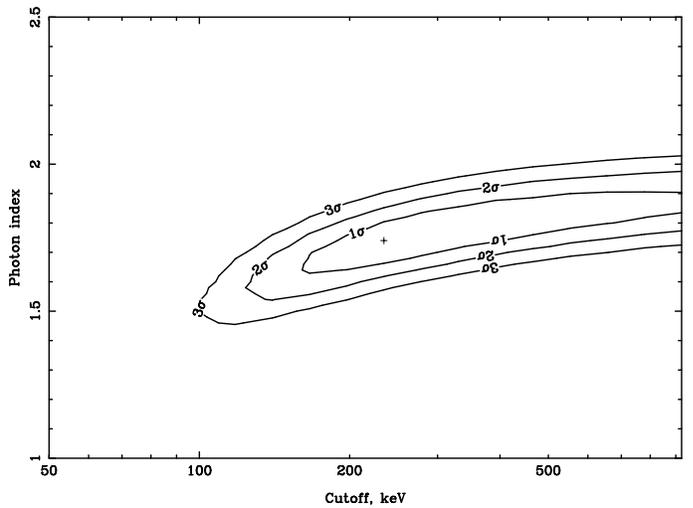}
\caption{Same as Fig.~\ref{fig:contour_l41_43.5}, but for the
cumulative SED of high-luminosity AGN ($\lhx>10^{43.5}$~erg~s$^{-1}$). 
}
\label{fig:contour_lg43.5}
\end{figure}

The above model provides an excellent fit both to the cumulative
SED of low-luminosity AGN and to that of high-luminosity ones. The
best-fit parameter values are given in Table~\ref{tab:bestfit} and the
associated confidence regions for the two main parameters $\Gamma$
and $E_f$ are shown in Figs.~\ref{fig:contour_l41_43.5} and
\ref{fig:contour_lg43.5}. The best-fit models are shown in 
Figs.~\ref{fig:bestfit_l41_43.5} and \ref{fig:bestfit_lg43.5}, as well
as the contributions of unobscured ($\log\nh<22$), obscured ($22<
\log\nh<24$) and heavily obscured ($\log\nh>24$) AGN to the total SEDs. 

In summary, the best-fit power-law photon index is found to be $\sim
1.7$, as expected of AGN (see e.g. \citealt{reynolds97,turetal97}). In
both studied cases, the model prefers to have a high-energy cutoff
with $E_f\sim 100$--300~keV, while its detection is only marginal
($<2\sigma$) in both cases. Previously, similar values of the cutoff
energy have been inferred from individual spectra of several brightest Seyfert
galaxies (in particular NGC~4151, \citealt{jouetal92}) selected without 
regard to their luminosities or space densities, as well as upon averaging 
over such spectra (e.g. \citealt{zdzetal95,peretal02,moletal06}).  

\begin{table*}
\caption{Results of the modeling of the cumulative SEDs of low- and
  high-luminosity AGN
\label{tab:bestfit}
}
\begin{center}

\begin{tabular}{cccccc}
\hline
\hline
\multicolumn{1}{c}{$L_{\rm hx}$} &
\multicolumn{1}{c}{$\Gamma$} &
\multicolumn{1}{c}{$E_f$} &
\multicolumn{1}{c}{$A$ ({\rm INTEGRAL})} &
\multicolumn{1}{c}{$A$ ({\rm RXTE})} &
\multicolumn{1}{c}{$\chi^2/$d.o.f.} \\

\multicolumn{1}{c}{erg~s$^{-1}$} &     
\multicolumn{1}{c}{} &     
\multicolumn{1}{c}{keV} &     
\multicolumn{2}{c}{$10^{38}$~erg~s$^{-1}$Mpc$^{-3}$} &
\multicolumn{1}{c}{} \\   
\hline
%
% $10^{41}$--$10^{43.5}$ & $1.64\pm0.17$ & $124_{-38}^{+134}$ & 3.0\pm1.4$ &
% $1.7\pm0.4 & 3.7/5 \\  divide INT by 1.1 and multiply RXTE by 1.43*1.1
%
$10^{41}$--$10^{43.5}$ & $1.64\pm0.17$ & $124_{-38}^{+134}$ & 
$2.7\pm1.3\pm0.5$ & $2.7\pm0.6\pm 0.6$ & 3.7/5 \\ 
%
% $>10^{43.5}$ & $1.74\pm0.08$ & $258_{-79}^{+277}$ & $0.55\pm0.12$ &
% $0.31\pm0.04$ & 3.1/5 \\ multiply RXTE by 1.43
%
$>10^{43.5}$ & $1.74\pm0.08$ & $258_{-79}^{+277}$ & $0.55\pm0.12\pm0.12$ &
$0.44\pm0.06\pm0.07$ & 3.1/5 \\ 
\hline
\end{tabular}

\end{center}

\end{table*}

We note the excellent agreement of the derived normalizations of the
low-energy (RXTE) and high-energy (INTEGRAL) parts of the cumulative
SED of low-luminosity AGN, i.e. the whole 3--300~keV spectrum can be
well fit by a model using a single normalization. Also, the $\sim
20$\% difference in the normalizations of the RXTE and 
INTEGRAL parts of the cumulative SED of high-luminosity AGN is
not statistically significant given the uncertainties
associated with the photon counting statistics and the finite size of the
AGN samples (see the first and second sets of uncertainties for the $A$
values in Table~\ref{tab:bestfit}). The total 1$\sigma$ uncertainties
are shown as shaded regions in Figs.~\ref{fig:bestfit_l41_43.5} and
\ref{fig:bestfit_lg43.5} and one can see that the INTEGRAL and RXTE
data are in good agreement with each other. Furthermore, part of
the apparent mismatch between the RXTE and INTEGRAL parts of the cumulative
SED of high-luminosity AGN may be caused by possible small 
inhomogeneities in the distribution of matter on the 100--200~Mpc scales 
(see \S\ref{s:lss}).

Apart from the relatively simple model given by
equation~(\ref{eq:model}), we also fit the measured SEDs by
more complicated models. In particular we tried adding a Compton reflection
component. Since this did not lead to a significant improvement of the
quality of the fit nor did it provide interesting constraints on the amplitude
of the reflection component (due to its strong correlation with the
power-law slope and the position of the high-energy cutoff), we base
all of the remaining discussion on our model of a sum of absorbed and
unabsorbed power-law spectra with a high-energy cutoff.

\subsection{Cumulative SED of the local AGN population}
\label{s:final}

We have seen that the shapes of the summed SEDs of low- and
high-luminosity AGN are similar to each other except that the
spectrum is harder below 20~keV in the former case due to
the larger fraction of obscured sources at low luminosities
($\lhx< 10^{43.5}$~erg~s$^{-1}$). We can now finally determine the SED of 
the hard X-ray volume emissivity
of all local AGN with luminosities over $10^{41}$~erg~s$^{-1}$.

To this end, we first slightly rescale the best-fit spectral models
presented above. Specifically, we define the amplitudes of the
low- and high-luminosity spectral components to be the
averages of the corresponding INTEGRAL and RXTE values given in
Table~\ref{tab:bestfit}, so that 
\beq
A_{\rm low}= 2.7\times 10^{38}~{\rm erg}~{\rm s}^{-1}~{\rm
  Mpc}^{-3}
\label{eq:alow}
\eeq
and
\beq
A_{\rm high}= 0.49\times 10^{38}~{\rm erg}~{\rm s}^{-1}~{\rm
  Mpc}^{-3}.
\label{eq:ahigh}
\eeq

\begin{figure}
\centering
\includegraphics[width=\columnwidth]{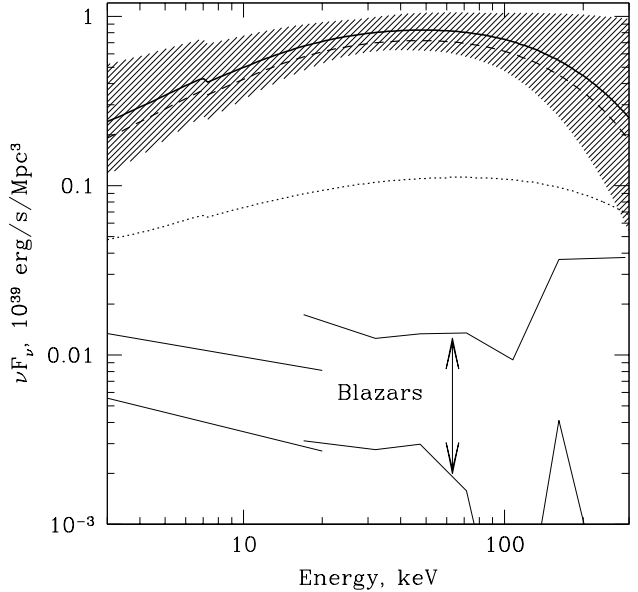}
\caption{Cumulative SED of local non-blazar AGN with luminosities over
$10^{41}$~erg~s$^{-1}$. The solid line is our best estimate of
this spectrum based on INTEGRAL and RXTE data. The shaded 
region represents the multifold of spectral models that are consistent
with the data to within 1$\sigma$ and includes the 20\%
uncertainty in the spectral normalization due to a combination of
systematic effects discussed in the text. The dashed and dotted
line show the contributions of  
low-luminosity ($10^{41}<\lhx<10^{43.5}$~erg~s$^{-1}$) and
high-luminosity  ($\lhx>10^{43.5}$~erg~s$^{-1}$) AGN,
respectively. The two solid curves at the bottom of the plot
demonstrate the contribution of blazars to the total local emissivity
at different energies, estimated based on INTEGRAL and RXTE data, with
the distance between the curves corresponding to the $1\sigma$ uncertainty. 
}
\label{fig:bestfit_total}
\end{figure}

In Fig.~\ref{fig:bestfit_total} we show by the solid line the sum of our low-
and high-luminosity best-fit models with the normalizations given by
equations~(\ref{eq:alow}) and (\ref{eq:ahigh}). This is our best
estimate of the cumulative SED of local AGN. The shaded 
region represents the multifold of spectral models that are consistent
with the INTEGRAL and RXTE data to within 1$\sigma$ and includes an
additional uncertainty ($\pm 20$\% around the average value) in the
spectral normalization, which is our estimate of a combination
of several uncertainties, which were already discussed in different
sections of the paper, namely associated with 1) the finite size of
the INTEGRAL and RXTE samples of AGN, 2) the local LSS (although we have 
taken it into account to a first approximation)
and 3) possible incompleteness of both AGN samples. The dashed and dotted
lines show the contributions of the low-luminosity
($10^{41}<\lhx<10^{43.5}$~erg~s$^{-1}$) and  high-luminosity
($\lhx>10^{43.5}$~erg~s$^{-1}$) AGN, respectively. We note that the
latter contribute just $\sim 15$\% to the collective SED of local AGN.

We finally note that blazars are not expected to contribute more than
several per cent to the local hard X-ray emissivity at different
energies below $\sim 200$~keV (see Fig.~\ref{fig:bestfit_total}). We
made this estimate using the same $1/V_{\rm max}$ method that we used
to derive the cumulative SED of non-blazar AGN. Although in
making this estimate we did not exclude high-redshift flat-spectrum
radio quasars present in the INTEGRAL and RXTE catalogs, it is
essentially based on the 2 and 9 nearby ($z<0.1$) BL Lacs detected by INTEGRAL
and RXTE, respectively. We also note that the contribution of blazars
may well be much more significant in gamma-rays ($E\ga
200$~keV), since emission from blazars can extend to very high
energies, in contrast to normal Seyfert galaxies and quasars.

\section{Comparison with the cosmic X-ray background}
\label{s:cxb}

In the previous section we derived the cumulative hard X-ray SED 
of local AGN. Specifically, we determined its normalization,
i.e. the volume hard X-ray emissivity of AGN, and shape, which
was shown to be well described as a sum of absorbed and unabsorbed
power-law spectra with a high-energy cutoff. It is interesting to put these
results into the broader perspective of the cosmic history of growth
of massive black holes and the origin of the cosmic X-ray background. 

To this end, suppose that the shape of the cumulative AGN SED does not 
evolve with redshift, which implies a constant relative contribution of 
obscured sources, while its normalization, i.e. the luminosity density of AGN,
does experience evolution. The collective emission of AGN from all redshifts 
observed at $z=0$ will then have the spectrum (for a flat cosmoslogy, see e.g.
\citealt{sazetal04}) 
\beq
I(E)=\frac{c}{4\pi H_0}\int_0^\infty\frac{\epsilon(z) S((1+z)E)}
{(1+z)\left[\Omega_{\rm m}(1+z)^3+\Omega_\Lambda\right]^{1/2}}dz,
\label{eq:bgr}
\eeq
where $S(E)$ is the cumulative SED of local AGN   
(Fig.~\ref{fig:bestfit_total}) and the function $\epsilon(z)$ describes the 
evolution of the AGN luminosity density.

Following \cite{sazetal07}, we assume that the hard X-ray luminosity density 
of AGN evolves similarly to the luminosity density of AGN with 
$\lhx\ga 10^{42}$~erg~s$^{-1}$ in the rest-frame 2--8~keV energy band as 
reported by \cite{baretal05}, specifically that
$\epsilon(z)$ is bound between two limiting functions:
\beqa
\epsilon_1(z)\propto 
\left\{
\begin{array}{ll}
(1+z)^{3.2},\,\,& z\le 1\\
\epsilon_1(1)/z,\,\,& z>1
\end{array}
\right.
\label{eq:e1}
\eeqa
and
\beqa
\epsilon_2(z)\propto
\left\{
\begin{array}{ll}
(1+z)^{3.2},\,\,& z\le 1\\
\epsilon_2(1),\,\,& z>1.
\end{array}
\right.
\label{eq:e2}
\eeqa
This formulation reflects the significant uncertainty in our
knowledge of the AGN evolution at $z\ga 1$, whereas the rapid, 
approximately power-law evolution between $z=0$ and $z\sim 1$ is  
established relatively well \citep{uedetal03,baretal05,lafetal05}.

\begin{figure}
\centering
\includegraphics[width=\columnwidth]{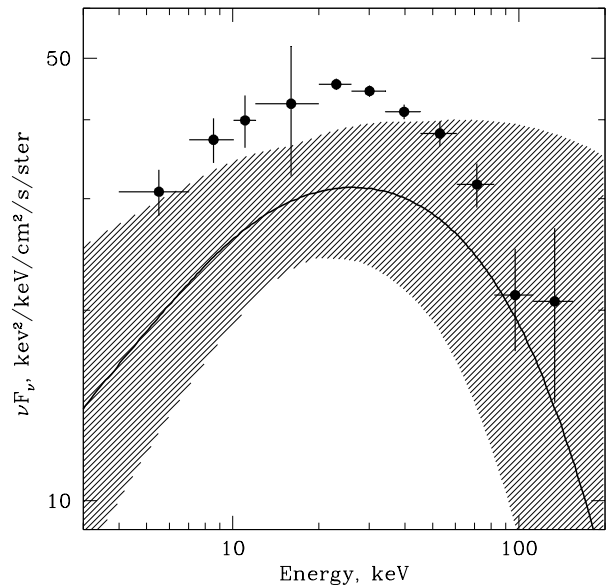}
\caption{Result of convolution of the cumulative SED of local AGN 
(Fig.~\ref{fig:bestfit_total}) with the redshift evolution function of the AGN
luminosity density described by equation~(\ref{eq:e1}). The solid line
is our best estimate of this redshift-integrated AGN SED. The
shaded region represents the $1\sigma$ uncertainty in the spectral shape 
combined with the 20\% uncertainty in the normalization. The points with 
error bars show the CXB spectrum measured with the JET-X, IBIS/ISGRI and SPI 
instruments on INTEGRAL \citep{chuetal07}.
}
\label{fig:cxb_spectrum_pl1}
\end{figure}

\begin{figure}
\centering
\includegraphics[width=\columnwidth]{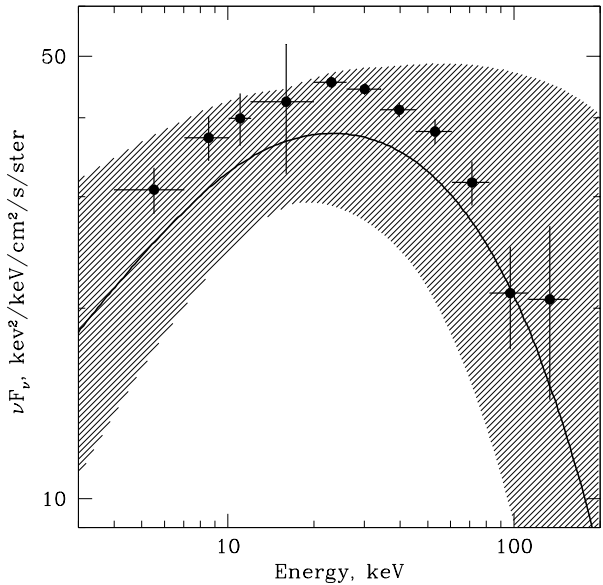}
\caption{Same as Fig.~\ref{fig:cxb_spectrum}, but for the scenario of
  evolution given by equation~(\ref{eq:e2}).
}
\label{fig:cxb_spectrum}
\end{figure}

We now use equation~(\ref{eq:bgr}) to convolve our locally determined
composite AGN SED (Fig.~\ref{fig:bestfit_total}) with the two
limiting evolution functions $\epsilon_1$ and $\epsilon_2$ (up to 
$z_{\rm max}=6$). The resulting SEDs, representing the collective emission of 
AGN from all redshifts, are shown in Fig.~\ref{fig:cxb_spectrum_pl1} and 
Fig.~\ref{fig:cxb_spectrum}, respectively. The shaded uncertainty regions
in these figures have been derived by propagating the uncertainties 
associated with the local AGN SED shown in Fig.~\ref{fig:bestfit_total}, 
including the $1\sigma$ uncertainty in the spectral shape and the estimated 
20\% uncertainty in the volume emissivity of local AGN.  

We see that within the fairly large uncertainties the shape of
the predicted redshift-integrated AGN SED is in good agreement with 
that of the CXB for both limiting scenarios of AGN evolution at $z>1$ and thus
for the true evolution constrained between $\epsilon_1(z)$ and 
$\epsilon_2(z)$. As concerns the normalization, depending on the character of 
evolution at $z>1$ our model is between being consistent with the observed CXB
intensity (in the case of $\epsilon_2(z)$) and underestimating it by
some 30\% (in the case of $\epsilon_1(z)$). We also note that by 
adopting the functional form $\epsilon(z)=(1+z)^{3.2}$ at $z<1$ we 
essentially fixed the ratio of the AGN luminosities densities at $z=1$ and 
$z=0$ at 9.2. In reality, this ratio is known with accuracy $\sim 20$\% 
\citealt{uedetal03,baretal05,lafetal05} and this uncertainty of course 
translates into a similar uncertainty in our predicted spectrum. The 
calculated SEDs are compared in Fig.~\ref{fig:cxb_spectrum_pl1} and 
\ref{fig:cxb_spectrum} with the CXB spectrum measured by INTEGRAL 
\citep{chuetal07}.

We note that if the simple scenario of evolution considered here is 
correct, we may expect the calculated SED to somewhat understimate the 
CXB intensity. Indeed, the calculation above rested on our estimate of the 
local AGN emissivity, which was based on non-blazar AGN with luminosities 
higher than $10^{41}$~erg~s$^{-1}$. An additional significant  
contribution to the local emissivity is expected to come from AGN with still 
lower luminosities ($\lhx<10^{41}$~erg~s$^{-1}$) \citep{elvetal84} and  
from blazars (see \S\ref{s:final} above). In fact, we recently used RXTE 
observations to study the correlation of large-scale variations of the CXB 
intensity with the spatial distribution of galaxies in the local Universe, 
which allowed us to estimate the combined emissivity of all X-ray sources 
(AGN, star-forming and normal galaxies, and clusters of galaxies etc) in the 
local Universe at $(9\pm 4)\times 10^{38}$~erg~s$^{-1}$~Mpc$^{-1}$ in the
2--10~keV energy band \citep{revetal08}. On the other hand, by
integrating the luminosity function of nearby AGN measured by RXTE
\citep{sazrev04} and applying the corrections described in
\S\ref{s:rxte}, we find that the combined emissivity of AGN with
$\lhx>10^{41}$~erg~s$^{-1}$ is $\sim 6\times
10^{38}$~erg~s$^{-1}$~Mpc$^{-1}$. Therefore, low-luminosity
($\lhx<10^{41}$~erg~s $^{-1}$) X-ray sources may indeed contribute
$\sim 30$\% or even more to the total local X-ray emissivity (at
least in the 2--10~keV energy band and probably somewhat less at
higher energies taking into account that clusters of galaxies
and normal galaxies produce softer X-ray emission than AGN).

We also point out that our calculation was based on the assumption that 
the evolution of the total luminosity density of AGN follows the observed 
evolution of the 2--8~keV luminosity density of AGN with 
$\lhx\ga 10^{42}$~erg~s$^{-1}$ \citep{baretal05}. The inclusion of 
lower-luminosity ($\lhx\la 10^{42}$~erg~s$^{-1}$) AGN, which 
are not yet accessible to observation at $z\ga 1$ and whose abundance is thus
poorly known, could also modify the result of our calculation.

We finally note the well-known fact that if the luminosity density of AGN were
the same at high redshifts as it is at $z=0$, only $\sim 20$\% of the CXB 
could be explained. 

\section{Discussion and conclusions}
\label{s:summary}

In this paper we used the INTEGRAL and RXTE all-sky hard X-ray surveys
to calculate the spectral energy distribution of the collective
emission of local AGN in the broad range 3--300~keV, properly
taking into account the relative contributions of AGN with different
luminosities and absorption column densities.

We first performed stacking spectral analyses separately for AGN with
low ($10^{41}<\lhx<10^{43.5}$~erg~s$^{-1}$) and high 
($\lhx>10^{43.5}$~erg~s$^{-1}$) luminosities and found that both stacked SEDs 
are consistent with having the same shape at energies above 20~keV -- in 
particular, a high-energy cutoff is marginally detected above 
$\sim $100--200~keV, whereas the cumulative SED of 
low-luminosity AGN is harder at energies below 20~keV due to the
larger fraction of obscured AGN compared to higher luminosities. We
then summed up our best-fit models for the stacked spectra of low- and 
high-luminosity AGN to obtain the collective SED of all local 
non-blazar AGN with $\lhx>10^{41}$~erg~s$^{-1}$ and estimated the associated 
statistical and systematic uncertainties. 

As a first attempt to apply the derived cumulative SED of local AGN to 
studying the history of black hole growth in the Universe, we demonstrated 
that this SED is consistent with the cosmic X-ray background spectrum if both 
the spectral shape of the collective hard X-ray emission of AGN and the 
relative fraction of obscured AGN remain constant with redshift while the 
total AGN luminosity density undergoes strong evolution between $z\sim 1$ and 
$z=0$. 

This simple model underpredicts the observed CXB intensity by $\sim 0$--30\% 
depending on the loosely constrained evolution of AGN at $z>1$ (and to a
lesser degree on the evolution at $z<1$). As we discussed in 
\S\ref{s:cxb}, the missing CXB flux can also possibly be attributed to blazars
 and to low-luminosity AGN ($\lhx\la 10^{41}$~erg~s$^{-1}$ at $z=0$ and 
$\lhx\la 10^{42}$~erg~s$^{-1}$ at $z\sim 1$--2), not included in our model, 
since such objects are known to be abundant and energetically 
important at least in the local Universe. In this connection we note 
that there may be an increased fraction of Compton thick 
($\nh\ga 10^{24}$~cm$^{-2}$) objects among such low-luminosity AGN compared 
to the $\sim 15$\% fraction at $\lhx\ga 10^{41}$~erg~s$^{-1}$ observed
by INTEGRAL. Indeed, some studies suggest that the intrinsic fraction of 
Compton thick AGN in the local Universe may be as high as $\sim 50$\% 
(e.g. \citealt{risetal99,guaetal05}) and in the most obscured of these objects
even the hard X-ray flux will be strongly depressed.

Turning the above argument around, we can conclude that AGN with 
$\lhx\la 10^{42}$~erg~s$^{-1}$, which are undetectable even in the deepest 
extragalactic surveys with Chandra and XMM-Newton, cannot provide a dominant
contribution to the X-ray emissivity of the Universe at redshifts $z\sim 1$--2,
otherwise they would produce too much X-ray background. 

How does this assumed scenario of AGN evolution compare with
observational data obtained by X-ray surveys at energies 
below $\sim 8$~keV? First of all, a number of studies have shown
that AGN undergo approximately pure luminosity evolution
between $z\sim 1$ and $z=0$ \citep{uedetal03,baretal05}, which means
that the AGN luminosity function has shifted to lower luminosities
by nearly an order of magnitude. More fundamentally, it has been
demonstrated that black hole growth and galaxy formation move steadily and in 
parallel to lower mass scales since $z\sim 2$ up to $z=0$ 
(e.g. \citealt{hecetal04}). 

This picture of ``cosmic downsizing''  or `` antihierarchical evolution'' is
consistent with the scenario considered in this paper and in fact in our 
calculations we adopted the functional form of the AGN luminosity density 
evolution between $z=0$ and $z\sim 1$ from the recent X-ray surveys. As 
concerns the relative contribution of obscured AGN to the cumulative hard 
X-ray emissivity at a given redshift, which is constant in our model, 
observations seem to indicate that there is indeed no significant evolution 
of this fraction between $z=0$ and $z\sim 1$ \citep{uedetal03}. Therefore, all
the available AGN observations at $z\la 1$ appear to well fit in the simple 
scenario of evolution considered in this paper.

However, X-ray observations also suggest that the character of AGN
evolution changes at higher redshifts ($z\ga 1$). Namely, the
observed AGN evolution is much better described by a luminosity
dependent density evolution model than by a pure luminosity
evolution one \citep{uedetal03,lafetal05,giletal07}. Also the relative
contribution of obscured sources to the total AGN luminosity density may have
undergone substantial evolution at $z\ga 1$,
although this remains a very controversial issue 
\citep{lafetal05,akyetal06,tozetal06,treurr06}. Although the different 
character of AGN evolution at high redshifts with respect to $z<1$
suggests that our simple model needs to be modified, such a revision
would remain almost unnoticed in view of the fairly large
uncertainties in the cumulative SED of local AGN. Indeed, as has been shown 
in \S\ref{s:cxb}, the current uncertainty in the AGN evolution at $z>1$ leads 
to less than $\sim 30$\% uncertainty in the predicted CXB spectrum, which is 
of the same order as the uncertainties in the cumulative SED of local AGN 
obtained from INTEGRAL and RXTE data. 

We also note that in reality the hard X-ray spectra of
distant quasars may somewhat differ from those of the nearby
Seyfert galaxies that make up our local cumulative SED, since,
although AGN spectra are not expected to directly depend
on cosmological redshift, they can depend on physical parameters of the 
accretion disk around the central massive black hole, determined by
the black hole mass, accretion rate \citep{shasun76} and
spin. Observations with future hard X-ray telescopes will permit
direct tests of whether the high-energy spectra 
of quasars are similar to those of local AGN or not. 

The main result of this work is that for the first time a direct
comparison has been made between the cumulative hard X-ray SED of the local 
AGN population and the CXB spectrum, which
demonstrated that the commonly accepted paradigm of the CXB being a
superposition of AGN is generally correct. Improved measurements of the
cumulative SED and evolution of AGN by current and
future X-ray and hard X-ray astronomy missions will make it possible to obtain
tighter constraints on the cosmic history of black hole growth and the
AGN unification paradigm.

\smallskip
\noindent {\sl Acknowledgments} This work was partially supported by the
DFG-Schwerpunktprogramme (SPP 1177). We thank the referee for helpful
remarks on the paper. INTEGRAL is an ESA project funded by ESA member
states (especially the PI countries: Denmark, France, Germany, Italy,
Spain, Switzerland), Czech Republic and Poland, and with the
participation of Russia and the USA. 

%%%%%%%%%%%%%%%%%%%%%%%%%%%%%%%%%%%%%%%%%%%%%%%%%%%%%%%%%%%%%%%%%%%

\end{document}